\documentclass[fdp]{w-art}

\usepackage{amsmath}
\usepackage{amsfonts}
\usepackage{amssymb}
\usepackage{graphicx}
\usepackage{citesort}
\usepackage{mathrsfs}

\def\S{{\mathbb S}}

\newcommand{\alg}[1]{\mathfrak{#1}}
\newcommand{\su}{\alg{su}}

\def\ads{{\rm AdS}_5\times {\rm S}^5}

\begin{document}


\title{Yangian Symmetry, S-Matrices and Bethe Ansatz for the $\ads$ Superstring}

\author[M de Leeuw]{M. de Leeuw\inst{1,}
  \footnote{E-mail:~\textsf{M.deLeeuw@uu.nl}}}
\address[\inst{1}]{{\it Institute for Theoretical
Physics and Spinoza Institute,\\ Utrecht University, 3508 TD
Utrecht, The Netherlands}}

\begin{abstract}
We discuss the relation between the recently derived bound state
S-matrices for the $\ads$ superstring and Yangian symmetry. We
will study the relation between this Yangian symmetry and the
Bethe ansatz. In particular we can use it to derive the Bethe
equations for bound states.
\end{abstract}

\maketitle

\section{Introduction and motivation}

The AdS/CFT correspondence \cite{Maldacena:1997re} has been the
subject of intensive research. One of the best studied examples of
this correspondence is $\mathcal{N}=4$ SYM gauge theory which is
conjectured to be dual to superstring theory on $\ads$. However,
testing or proving the duality for these concrete models poses a
very challenging problem.

A breakthrough in the understanding of this duality was the
discovery of integrable structures. Integrable structures were
found both in $\mathcal{N}=4$ SYM  and in the $\ads$ string sigma
model \cite{Minahan:2002ve,Bena:2003wd}. Because of the huge
number of symmetries, integrable theories are special. For
example, in scattering processes for integrable theories, the set
of particle momenta is conserved and scattering processes always
factorize into a sequence of two-body interactions. This implies
that in such theories the scattering information is encoded in the
two-body S-matrix. However, a full proof of integrability for both
$\mathcal{N}=4$ SYM and superstrings on $\ads$ is currently still
lacking.

Nevertheless, one can assume integrability and try to exploit its
consequences to solve the theory. Then one can compare the
obtained results to explicit calculations afterwards. This is the
current approach and it proved to be quite fruitful. By assuming
integrability one can make use of the S-matrix approach. This
leads to the conjecture of the ``all-loop" Bethe equations
describing the gauge theory asymptotic spectrum
\cite{Beisert:2005fw}. Also for the $\ads$ superstring, a Bethe
ansatz for the $\su(2)$ sector was proposed
\cite{Arutyunov:2004vx}. Actually, in both theories, exact two
body S-matrices were found \cite{Beisert:2005tm,Arutyunov:2006yd}.
These enabled the use of the Bethe ansatz \cite{Beisert:2005tm},
confirming the conjectured Bethe equations for physical states.

More precisely, the two body S-matrix is almost completely fixed
by symmetry. Both the asymptotic spectrum of $\mathcal{N}=4$ super
Yang-Mills theory \cite{Beisert:2005tm} and the light-cone
Hamiltonian \cite{Arutyunov:2006ak} for the $\ads$ superstring
exhibit the same symmetry algebra; centrally extended
$\mathfrak{su}(2|2)$. The requirement that the S-matrix is
invariant under this algebra determines it up to an overall phase
\cite{Beisert:2005tm} and the choice of representation basis
\cite{Arutyunov:2006yd}. In a suitable local scattering basis, the
S-matrix exhibits most properties of massive two-dimensional
integrable field theories, like crossing symmetry
\cite{Janik:2006dc} and the Yang-Baxter equation
\cite{Arutyunov:2006yd}. The overall (dressing) phase appears to
be a remarkable feature of the string S-matrix and it has been
studied intensely, see e.g. \cite{Beisert:2006ez}.

The string sigma model also contains an infinite number of bound
states \cite{Arutyunov:2007tc}. These bound states fall into short
(atypical) symmetric representations of the centrally extended
$\mathfrak{su}(2|2)$ algebra \cite{Dorey:2006dq}. Of course, these
states scatter via their own S-matrices. Recently a number of
these bound states S-matrices have been found
\cite{Arutyunov:2008zt,Bajnok:2008bm}. However, invariance under
centrally extended $\mathfrak{su}(2|2)$ is not enough to fix all
these S-matrices. One needs to impose the Yang-Baxter equation by
hand in order to completely fix them up to a phase.

The found asymptotic Bethe ans\"atze only describe the spectra in
the infinite volume limit. For a complete check of this particular
case of the AdS/CFT correspondence, the full spectra have to be
computed and compared. Away from the asymptotic region wrapping
interactions appear. One way to include these is L\"uscher's
perturbative approach \cite{Janik:2007wt}. In L\"uscher's approach
one deals with corrections coming from virtual particles that
propagate around the compact direction. These virtual particles
can be both fundamental particles and bound states. This method
has proven successful since it recently allowed the computation of
the full four-loop Konishi operator with wrapping interactions
\cite{Bajnok:2008bm}. The result coincided with the gauge theory
computation \cite{Fiamberti:2007rj}, providing an very non-trivial
check of the correspondence.

The approach by L\"uscher is closely related to the thermodynamic
Bethe ansatz (TBA). In the TBA one deals with finite size effects
by defining a mirror model \cite{Arutyunov:2007tc}. Finite size
effects in the original theory correspond to finite temperature
effects in the infinite volume for the mirror theory. Here again,
one needs to include all (physical) bound states. One of the
advantages of this method is that one can still use the asymptotic
Bethe ansatz.

In other words, knowledge of the bound states, their S-matrices
and the corresponding Bethe ans\"atze is crucial for a complete
understanding of finite size effects. In this proceedings, we will
give a short overview of the role of Yangian symmetry in this.
More specifically, we will sketch how Yangian symmetry enables one
to find the asymptotic Bethe equations for bound states
\cite{deLeeuw:2008ye}. This paper is organized as follows, first
we discuss centrally extended $\alg{su}(2|2)$, its Yangian and how
one can derive S-matrices from this. Then we will briefly explain
how the Bethe Ansatz works and its relation to Yangian symmetry.

\section{The Yangian of $\mathfrak{su}(2|2)$ and S-Matrices}

In this section we will briefly discuss the Yangian of
$\mathfrak{su}(2|2)$ and bound state representations. For more
details see e.g \cite{deLeeuw:2008ye} and references therein.

\subsection{The algebra in superspace}

The superalgebra $\mathfrak{su}(2|2)$ is the symmetry algebra of
the $\ads$ superstring and it also is the symmetry algebra of the
spin chain connected to $\mathcal{N}=4$ SYM. The algebra consists
of bosonic generators $\mathbb{R},\mathbb{L}$, generating two
copies of $\mathfrak{su}(2)$, supersymmetry generators
$\mathbb{Q},\mathbb{G}$ and central charges
$\mathbb{H},\mathbb{C},\mathbb{C}^{\dag}$. The non-trivial
commutation relations are given by
\begin{eqnarray}
\begin{array}{lll}
\ [\mathbb{L}_{a}^{\ b},\mathbb{J}_{c}] = \delta_{c}^{b}\mathbb{J}_{a}-\frac{1}{2}\delta_{a}^{b}\mathbb{J}_{c} &\qquad & \ [\mathbb{R}_{\alpha}^{\ \beta},\mathbb{J}_{\gamma}] = \delta_{\gamma}^{\beta}\mathbb{J}_{\alpha}-\frac{1}{2}\delta_{\alpha}^{\beta}\mathbb{J}_{\gamma}\\
\ [\mathbb{L}_{a}^{\ b},\mathbb{J}^{c}] = -\delta_{a}^{c}\mathbb{J}^{b}+\frac{1}{2}\delta_{a}^{b}\mathbb{J}^{c} &\qquad& \ [\mathbb{R}_{\alpha}^{\ \beta},\mathbb{J}^{\gamma}] = -\delta^{\gamma}_{\alpha}\mathbb{J}^{\beta}+\frac{1}{2}\delta_{\alpha}^{\beta}\mathbb{J}^{\gamma}\\
\ \{\mathbb{Q}_{\alpha}^{\ a},\mathbb{Q}_{\beta}^{\
b}\}=\epsilon_{\alpha\beta}\epsilon^{ab}\mathbb{C}&\qquad&\ \{\mathbb{G}^{\ \alpha}_{a},\mathbb{G}^{\ \beta}_{b}\}=\epsilon^{\alpha\beta}\epsilon_{ab}\mathbb{C}^{\dag}\\
\ \{\mathbb{Q}_{\alpha}^{a},\mathbb{G}^{\beta}_{b}\} =
\delta_{b}^{a}\mathbb{R}_{\alpha}^{\ \beta} +
\delta_{\alpha}^{\beta}\mathbb{L}_{b}^{\ a}
+\frac{1}{2}\delta_{b}^{a}\delta_{\alpha}^{\beta}\mathbb{H}.&&
\end{array}
\end{eqnarray}
The generators $\mathbb{Q},\mathbb{G}$ are conjugate. For the
gauge fixed $\ads$ superstring, $\mathbb{H}$ corresponds to the
light-cone Hamiltonian and the central charge $\mathbb{C}$ depends
on the world-sheet momentum in the following way $\mathbb{C} =  i
g \zeta (e^{ip}-1)$.

A convenient way to describe representations of bound states is
the superspace formalism \cite{Arutyunov:2008zt}. Consider the
vector space of analytic functions of two bosonic variables
$w_{a}$ and two fermionic variables $\theta_{\alpha}$. The
$\mathfrak{su}(2|2)$ representation that describes $\ell$-particle
bound states of the light-cone string theory on $\ads$ is $4\ell$
dimensional and is spanned by monomials of degree $\ell$. This
representation is called the atypical totally symmetric
representation, and it describes $\ell$-particle bound states
\cite{Arutyunov:2008zt}. In this representation, the algebra
generators are represented by differential operators, depending on
parameters, $a,b,c,d$
\begin{eqnarray}\label{eqn;AlgDiff}
\begin{array}{lll}
  \mathbb{L}_{a}^{\ b} = w_{a}\frac{\partial}{\partial w_{b}}-\frac{1}{2}\delta_{a}^{b}w_{c}\frac{\partial}{\partial w_{c}}, &\qquad& \mathbb{R}_{\alpha}^{\ \beta} = \theta_{\alpha}\frac{\partial}{\partial \theta_{\beta}}-\frac{1}{2}\delta_{\alpha}^{\beta}\theta_{\gamma}\frac{\partial}{\partial \theta_{\gamma}}, \\
  \mathbb{Q}_{\alpha}^{\ a} = a \theta_{\alpha}\frac{\partial}{\partial w_{a}}+b\epsilon^{ab}\epsilon_{\alpha\beta} w_{b}\frac{\partial}{\partial \theta_{\beta}}, &\qquad& \mathbb{G}_{a}^{\ \alpha} = d w_{a}\frac{\partial}{\partial \theta_{\alpha}}+c\epsilon_{ab}\epsilon^{\alpha\beta} \theta_{\beta}\frac{\partial}{\partial w_{b}}
\end{array}.
\end{eqnarray}
The central charges are given by
\begin{eqnarray}
\begin{array}{ll}
 \mathbb{C} = ab \left(w_{a}\frac{\partial}{\partial w_{a}}+\theta_{\alpha}\frac{\partial}{\partial
 \theta_{\alpha}}\right),&
 \mathbb{H}= (ad +bc)\left(w_{a}\frac{\partial}{\partial w_{a}}+\theta_{\alpha}\frac{\partial}{\partial
 \theta_{\alpha}}\right).
\end{array}
\end{eqnarray}
To form a representation, the parameters $a,b,c,d$ must satisfy
$ad-bc = 1$. One can also express these parameters in terms of
particle momentum $p$ and coupling $g$:
\begin{eqnarray}
\begin{array}{lllllll}
  a = \sqrt{\frac{g}{2\ell}}\eta, & \ \ & b = \sqrt{\frac{g}{2\ell}}
\frac{i\zeta}{\eta}\left(\frac{x^{+}}{x^{-}}-1\right), & \ \ & c =
-\sqrt{\frac{g}{2\ell}}\frac{\eta}{\zeta x^{+}}, & \ \ &
d=\sqrt{\frac{g}{2\ell}}\frac{x^{+}}{i\eta}\left(1-\frac{x^{-}}{x^{+}}\right),
\end{array}
\end{eqnarray}
where $x^{\pm}$ are related to the particle momentum and coupling
constant via
\begin{eqnarray}
x^{+} +
\frac{1}{x^{+}}-x^{-}-\frac{1}{x^{-}}=\frac{2i\ell}{g},\qquad
\frac{x^{+}}{x^{-}} = e^{ip}.
\end{eqnarray}
The parameters $\eta$ parameterize the scattering basis and are
given by
\begin{eqnarray}\label{eqn;ScatteringBasis}
\eta = e^{i\xi}\eta(p),\qquad \eta(p)=
e^{\frac{i}{4}p}\sqrt{ix^{-}-ix^{+}},\qquad \zeta = e^{2i\xi}.
\end{eqnarray}
The fundamental representation is obtained by taking $\ell=1$.

The double Yangian $DY(\mathfrak{g})$ of a (simple) Lie algebra
$\mathfrak{g}$ is a deformation of the universal enveloping
algebra $U(\mathfrak{g}[u,u^{-1}])$ of the loop algebra
$\mathfrak{g}[u,u^{-1}]$. The Yangian is generated by a tower of
generators $\mathbb{J}^{A}_{n},\ n\in\mathbb{Z}$ that satisfy the
commutation relations
\begin{eqnarray}
\ [\mathbb{J}^{A}_{m},\mathbb{J}^{B}_{n}  ]  = F^{AB}_{C}
\mathbb{J}^{C}_{m+n} + \mathcal{O}(\hbar),
\end{eqnarray}
where $F^{AB}_{C}$ are the structure constants of the Lie algebra
$\mathfrak{g}$. The level 0 generators $\mathbb{J}_{0}^{A}$ span
the Lie-algebra itself. We are interested in the Yangian of
centrally extended $\su(2|2)$. This Yangian can be supplied with a
coproduct structure \cite{Beisert:2007ds,Plefka:2006ze}. The
coproduct of the $\su(2|2)$ operators is given by:
\begin{eqnarray}\label{eqn;CoprodSymm}
\Delta(\mathbb{J}_{0}^{A}) =\mathbb{J}_{0}^{A}\otimes 1 +
1\otimes\mathbb{J}_{0}^{A}.
\end{eqnarray}
We refer to \cite{Beisert:2007ds} for explicit formulas for the
Yangian generators.

An important representation of the Yangian is the evaluation
representation. It consists of states $|u\rangle$, with action
$\mathbb{J}^{A}_{n}|u\rangle = u^{n}\mathbb{J}^{A}_{0}|u\rangle$.
In this representation the coproduct structure is fixed in terms
of the coproducts of $\mathbb{J}_0,\mathbb{J}_1$. We will work in
this representation and identify
$\mathbb{J}_1\equiv\hat{\mathbb{J}} = \frac{g}{2i}u\mathbb{J}$ for
the $\su(2|2)$ Yangian. One finds that $u$ is dependent on the
parameters $x^{\pm}$ via $u_{j} = x_{j}^{+}+\frac{1}{x_{j}^{+}} -
\frac{i\ell_{j}}{g}$.

\subsection{S-matrices and symmetry}

Requiring invariance under centrally extended $\alg{su}(2|2)$
proved to be enough to fix the fundamental S-matrix up to a phase
factor \cite{Beisert:2005tm,Arutyunov:2006yd}. Symmetry invariance
means that for any generator $\mathbb{J}^{A}$ from
$\alg{su}(2|2)$, the S-matrix should satisfy
\begin{eqnarray}\label{eqn;SymmProp1}
\S~\Delta(\mathbb{J}^{A})&=&\Delta^{op}(\mathbb{J}^{A})~\S,
\end{eqnarray}
where $\Delta^{op} = P \Delta$, with $P$ the graded permutation.

When one uses this to compute bound state S-matrices one finds
that invariance under centrally extended $\alg{su}(2|2)$ is no
longer sufficient. In the case when one scatters two 2-particle
bound states, the S-matrix is still dependent on an additional
parameter. This parameter can be fixed by insisting that the
S-matrix satisfies the Yang-Baxter equation
\cite{Arutyunov:2008zt}.

However, there is an alternative to the Yang-Baxter equation to
completely fix the S-matrix. It was shown that the fundamental
S-matrix actually has a larger symmetry group than just
$\alg{su}(2|2)$, namely the Yangian of $\alg{su}(2|2)$
\cite{Beisert:2007ds}. This symmetry can again be understood in
terms of coproducts of generators, i.e. the S-matrix satisfies
\begin{eqnarray}\label{eqn;SymmProp2}
\S~\Delta(\hat{\mathbb{J}}^{A})&=&\Delta^{op}(\hat{\mathbb{J}}^{A})~\S
\end{eqnarray}
Then is was also shown that the bound state S-matrices respect
Yangian symmetry \cite{deLeeuw:2008dp}. Moreover, it was found
that the bound state S-matrices were fixed by Yangian symmetry
without reference to the Yang-Baxter equation. Hence, bound state
S-matrices, up to an overall phase, are completely fixed by
invariance under the full Yangian symmetry rather than under
$\alg{su}(2|2)$ alone. It seems that the Yangian of centrally
extended $\alg{su}(2|2)$ is the fundamental symmetry group
underlying the scattering processes of the $\ads$ superstring.

\section{The Bethe Ansatz and Yangian Symmetry}

For integrable systems, the number of particles and the set of
momenta is conserved. Let us consider $K^{\mathrm{I}}$ excitations
with momenta $p_{1},\ldots, p_{K^{\mathrm{I}}}$. We are dealing
with closed strings and hence we need to impose periodicity. A way
to do this is by using the so-called Bethe ansatz. This gives
certain restrictions on the particle momenta, formulated in terms
of the Bethe equations.

\begin{figure}[t]
  \centering
\includegraphics{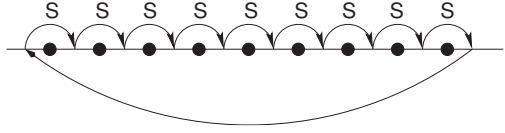}
  \caption{One particle is moved around the circle, scattering with the other particles.
  }\label{Fig;BA}
\end{figure}

In this approach one assumes that there are regions where the
particle coordinates $x_i$ are far apart in the sense that the
particles behave as free particles. In these asymptotic regions,
one can make a plane-wave ansatz for the wave function. Consider
permutations $\mathcal{P},\mathcal{Q}$, then the ansatz for the
wave function is of the form of a generalized Bethe ansatz
\cite{Yang:1967bm}
\begin{eqnarray}
|p_{1},\ldots,p_{K}\rangle = \sum_{\mathcal{P}}\int dx
\left\{A^{\mathcal{P}|\mathcal{Q}}_{a_{1}\ldots a_{K}}e^{i
p_{\mathcal{P}_i}x_{\mathcal{Q}_i}}\right\}\phi^{a_1}(x_1)\ldots
\phi^{a_K}(x_K),
\end{eqnarray}
where $\phi^{a_i}(x_i)$ creates a particle of type $a_i$ at
position $x_i$. This just corresponds making a linear combination
of plane waves for each ordering of the positions and momenta of
the particles. The interactions are described by the S-matrix and
they allow particles to cross the various regions and in this way
relate the coefficients
\begin{eqnarray}\label{eqn;compat}
A^{\mathcal{P}|\mathcal{Q}} &=& \S_{i,j}
A^{\mathcal{P}'|\mathcal{Q}'},
\end{eqnarray}
where the regions $\mathcal{P}|\mathcal{Q}$ and
$\mathcal{P}'|\mathcal{Q}'$ differ by permuting particles $i,j$.
Therefore, the scattering data give relations on the coefficients
from adjacent free regions. The Bethe equations are now of the
form
\begin{eqnarray}\label{eqn;BAE}
\S_{k k-1}\ldots \S_{k K^{\rm{I}}}\S_{k 1}\ldots
\S_{kk+1}A^{\mathcal{P}|\mathcal{Q}} =
e^{ip_{k}L}A^{\mathcal{P}|\mathcal{Q}}.
\end{eqnarray}
These equations can be interpreted in the following way. When a
particle is moved around the circle, it meets the other particles
and it scatters with them, see Figure \ref{Fig;BA}. When the
particle has moved around, the system should be unchanged up to a
phase factor.
The coordinate Bethe ansatz now proceeds by making an ansatz for
the coefficients $A^{\mathcal{P}|\mathcal{Q}}$ in such a way that
they solve (\ref{eqn;compat}) and then reading off the Bethe
equations for the particle momenta.

One first defines a vacuum $
A_{1\ldots1}^{\mathcal{P}|\mathcal{Q}}:=|0\rangle$. We restrict to
the undressed S-matrix, i.e. $\S|0\rangle = |0\rangle$. The next
thing to consider is a fermion inserted in this vacuum and treat
is as an excitation. The coefficients describing these states will
depend on the particle momentum and one must check whether this
construction respects (\ref{eqn;compat}). It turns out that this
gives some restrictions on the coefficients, which can be solved
from this \cite{Beisert:2005tm}. The problem becomes more involved
upon inserting multiple fermions. Here a new S-matrix
$\S^{\mathrm{II}}$ is introduced. Plugging this into
(\ref{eqn;compat}) again allows one to find explicit (unique)
solutions for the coefficients, including $\S^{\mathrm{II}}$. We
can deal with $\S^{\mathrm{II}}$ by using an additional Bethe
ansatz. This is called nesting. One can repeat the above procedure
to deal with this. For details we refer to \cite{Beisert:2005tm}.
This procedure results in the well-known Bethe equations
describing the asymptotic spectrum of the $\ads$ superstring:
\begin{eqnarray}
e^{ip_{k}L}&=& \prod_{l=1,l\neq k
}^{K^{\mathrm{I}}}\left[S_{0}(p_{k},p_{l})\frac{x_k^+-x_l^-}{x_k^--x_l^+}\sqrt{\frac{x_l^+x_k^-}{x_l^-x_k^+}}\right]^2\prod_{\alpha=1}^{2}\prod_{l=1}^{K_{(\alpha)}^{\mathrm{II}}}\frac{{x_{k}^{-}-y^{(\alpha)}_{l}}}{x_{k}^{+}-y^{(\alpha)}_{l}}\sqrt{\frac{x^+_k}{x^-_k}}\nonumber \\
1&=&\prod_{l=1}^{K^{\mathrm{I}}}\frac{y^{(\alpha)}_{k}-x^{+}_{l}}{y^{(\alpha)}_{k}-x^{-}_{l}}\sqrt{\frac{x^-_k}{x^+_k}}
\prod_{l=1}^{K_{(\alpha)}^{\mathrm{III}}}\frac{y_{k}^{(\alpha)}+1/y_{k}^{(\alpha)}-w_{l}^{(\alpha)}+i/g}{y_{k}^{(\alpha)}+1/y_{k}^{(\alpha)}-w_{l}^{(\alpha)}-i/g}\\
1&=&\prod_{l=1}^{K_{(\alpha)}^{\mathrm{II}}}\frac{w_{k}^{(\alpha)}-y_{k}^{(\alpha)}-1/y_{k}^{(\alpha)}+i/g}{w_{k}^{(\alpha)}-y_{k}^{(\alpha)}-1/y_{k}^{(\alpha)}-i/g}
\prod_{l\neq
k}^{K_{(\alpha)}^{\mathrm{III}}}\frac{w_{k}^{(\alpha)}-w_{l}^{(\alpha)}-2i/g}{w_{k}^{(\alpha)}-w_{l}^{\alpha}+2i/g}\nonumber,
\end{eqnarray}
with $\alpha=1,2$ and $S_{0}(p_{k},p_{l})$ the overall phase of
the S-matrix.

It is important to note that in this derivation the explicit form
of the S-matrix was used. However, not all bound state S-matrices
are known and hence the above procedure cannot be
straightforwardly applied to bound states.

\subsection{Bethe Ansatz and Yangian Symmetry}

The Bethe ansatz can be reformulated by considering coproducts of
(Yangian) symmetry generators. This formulation allows us to find
solutions to (\ref{eqn;compat}) without knowing the explicit form
of the bound state S-matrix. For details and explicit formulae, we
refer to \cite{deLeeuw:2008ye}.

Again one starts by defining a vacuum. The first non-trivial step
is to consider a fermion in this vacuum. The remarkable fact is
that, when restricted to two sites, one can find functions
$K_0,K_1$ such that one can write this coefficient as
\begin{eqnarray}
\left(K_{0}(p_1,p_2)\Delta\mathbb{Q}^{1}_{\alpha}+K_{1}(p_1,p_2)\Delta\hat{\mathbb{Q}}^{1}_{\alpha}\right)|0\rangle.
\end{eqnarray}
The invariance of the S-matrix under Yangian symmetry implies that
\begin{eqnarray}
\S |\alpha\rangle
=\left(K_{0}(p_1,p_2)\Delta^{op}\mathbb{Q}^{1}_{\alpha}+K_{1}(p_1,p_2)\Delta^{op}\hat{\mathbb{Q}}^{1}_{\alpha}\right)|0\rangle,
\end{eqnarray}
since $\S|0\rangle = |0\rangle$. However, this means that
(\ref{eqn;compat}) corresponds to requiring that $K_0$ and $K_1$
are symmetric under interchanging $p_1 \leftrightarrow p_2$. Note
that this is solely based on symmetry algebra arguments and is
valid for \textit{any} bound state number. In other words,
(\ref{eqn;compat}) which appears in the Bethe ansatz proves to be
closely related to the symmetry properties of the S-matrix. The
symmetry properties of $K_0,K_1$ can then be used to extract
information about the Bethe equations.

A similar treatment can be done for two fermions. This again gives
rise to solutions for the factors that appear in the Bethe
equations. In particular one finds that the auxiliary S-matrix,
$\S^{\mathrm{II}}$ remains unchanged. This completely fixes the
Bethe equations.

In conclusion, by making use of coproducts and Yangian symmetry,
we have found a way, independent of the explicit form of the
S-matrix, to find the Bethe equations of string bound states. They
are explicitly given by
\begin{eqnarray}
e^{ip_{k}(L+K^{\mathrm{I}}-\frac{K_{1}^{\mathrm{II}}}{2}-\frac{K_{2}^{\mathrm{II}}}{2})}&=&
e^{iP}\prod_{l=1,l\neq k
}^{K^{\mathrm{I}}}S_{0}(p_{k},p_{l})^{2}\prod_{\alpha=1}^{2}\prod_{l=1}^{K_{(\alpha)}^{\mathrm{II}}}\frac{{x_{k}^{-}-y^{(\alpha)}_{l}}}{x_{k}^{+}-y^{(\alpha)}_{l}}\nonumber \\
1&=&e^{-i\frac{P}{2}}\prod_{l=1}^{K^{\mathrm{I}}}\frac{y^{(\alpha)}_{k}-x^{+}_{l}}{y^{(\alpha)}_{k}-x^{-}_{l}}
\prod_{l=1}^{K_{(\alpha)}^{\mathrm{III}}}\frac{y_{k}^{(\alpha)}+\frac{1}{y_{k}^{(\alpha)}}-w_{l}^{(\alpha)}+\frac{i}{g}}{y_{k}^{(\alpha)}+\frac{1}{y_{k}^{(\alpha)}}-w_{l}^{(\alpha)}-\frac{i}{g}}\\
1&=&\prod_{l=1}^{K_{(\alpha)}^{\mathrm{II}}}\frac{w_{k}^{(\alpha)}-y_{k}^{(\alpha)}-\frac{1}{y_{k}^{(\alpha)}}+\frac{i}{g}}{w_{k}^{(\alpha)}-y_{k}^{(\alpha)}-\frac{1}{y_{k}^{(\alpha)}}-\frac{i}{g}}
\prod_{l\neq
k}^{K_{(\alpha)}^{\mathrm{III}}}\frac{w_{k}^{(\alpha)}-w_{l}^{(\alpha)}-\frac{2i}{g}}{w_{k}^{(\alpha)}-w_{l}^{\alpha}+\frac{2i}{g}}\nonumber,
\end{eqnarray}
with
\begin{eqnarray}
x_{k}^{+}+\frac{1}{x_{k}^+}-x_{k}^{-}-\frac{1}{x_{k}^-} = \frac{2i
\ell_{k}}{g},\qquad \frac{x_{k}^+}{x_{k}^-} = e^{ip_k}, \qquad
\alpha=1,2.
\end{eqnarray}
Note that apart from the parameters $x^{\pm}$, the phase factor
$S_{0}(p_{k},p_{l})$ also implicitly depends on the bound states
number \cite{Arutyunov:2008zt}. The found Bethe equations coincide
with the Bethe equations one expects from a fusion procedure.

\section*{Acknowledgements}

I would like to thank the organizers of the 4-th EU RTN Workshop
in Varna for the opportunity to present this work. I am indebted
to G. Arutyunov, S. Frolov and A. Torrielli for valuable
discussions. This work was supported in part by the EU-RTN network
\textit{Constituents, Fundamental Forces and Symmetries of the
Universe} (MRTN-CT-2004-005104), by the INTAS contract 03-51-6346
and by the NWO grant 047017015.

\bibliographystyle{fdp}
\bibliography{LitRmat}

\end{document}